\documentclass[12pt,preprint]{aastex}

\shorttitle{Thermal stability of thin disk with winds}
\shortauthors{S.-L. Li \& M. C. Begelman}

\begin{document}


\title{Thermal stability of thin disk with magnetically driven winds}

\author{Shuang-Liang Li\altaffilmark{1,2}
and Mitchell C. Begelman\altaffilmark{2,3}}

\altaffiltext{1}{Key Laboratory for Research in Galaxies and
Cosmology, Shanghai Astronomical Observatory, Chinese Academy of
Sciences, 80 Nandan Road, Shanghai, 200030, China; lisl@shao.ac.cn}
\altaffiltext{2}{JILA, University of Colorado and National Institute of Standards and Technology, 440 UCB, Boulder, CO 80309, USA; mitch@jila.colorado.edu}
\altaffiltext{3}{Department of Astrophysical and Planetary Sciences, University of Colorado,
Boulder, CO 80309, USA}

\begin{abstract}

The absence of thermal instability in the high/soft state of black hole X-ray binaries, in disagreement with the standard thin disk theory, is a long-standing riddle for theoretical astronomers. We have tried to resolve this question by studying the thermal stability of a thin disk with magnetically driven winds in the $\dot{M}- \Sigma$ plane. It is found that disk winds can greatly decrease the disk temperature and thus help the disk become more stable at a given accretion rate. The critical accretion rate $\dot{M}_{\rm crit}$ corresponding to the thermal instability threshold is increased significantly in the presence of disk winds. For $\alpha=0.01$ and $B_{\rm \phi}=10B_{\rm _p}$, the disk is quite stable even for a very weak initial poloidal magnetic field [$\beta_{\rm p,0}\sim 2000, \beta_{\rm p}=(P_{\rm {gas}}+P_{\rm rad})/(B_{\rm p}^2/8\pi)$]. But when $B_{\rm \phi}=B_{\rm _p}$ or $B_{\rm \phi}=0.1B_{\rm _p}$, a somewhat stronger (but still  weak) field ($\beta_{\rm p,0}\sim 200$ or $\beta_{\rm p,0}\sim 20$) is required to make the disk stable. Nevertheless, despite the great increase of $\dot{M}_{\rm crit}$, the luminosity threshold corresponding to instability remains almost constant or even decreases slowly with increasing $\dot{M}_{\rm crit}$  due to the decrease of gas temperature. The advection and diffusion timescales of the large-scale magnetic field threading the disk are also investigated in this work. We find that the advection timescale can be smaller than the diffusion timescale in a disk with winds, because the disk winds take away most of the gravitational energy released in the disk, resulting in the decrease of the magnetic diffusivity $\eta$ and the increase of the diffusion timescale.

\end{abstract}


\section{INTRODUCTION}

According to the standard thin disk theory, the radiation pressure dominated inner region of a thin disk is both thermally and viscously unstable when the Eddington-scaled mass accretion rate is larger than a critical value \citep{s1973,l1974,s1976,p1978}, which corresponds to a few percent of the Eddington luminosity. However, the high/soft state of X-ray binaries appears quite stable on observation. \citet{g2004} found that black hole X-ray binaries with luminosities ranging from $0.01$ to $0.5$ $L_{\rm Edd}$ show little variability, which obviously conflicts with the accretion disk theory. Only one superluminous X-ray binary, GRS 1915+105, was found to possess the limit-cycle light curve expected to be  produced by thermal-viscous instability over the course of decades \citep{b1997}. The variability of GRS 1915+105 was inferred to be related to its high luminosity \citep{g2004}. But recently, \citet{a2011} reported another source, IGR J17091-3624, that seems to show variability similar to that of GRS 1915+105 at lower luminosity, which suggests that there may be other variables associated with limit-cycle behavior in X-ray binaries. Although the analogous limit-cycle in active galactic nuclei (AGN) is hard to observe directly due to its long timescale, some intermittent activity in young radio galaxies has been ascribed to thermal instability in the disk \citep{c2009,w2009}.

There are mainly two processes that can change the theoretical results. Firstly, if the disk viscous stress is proportional to the gas pressure instead of the total pressure, the disk will be stable \citep{s1981,s1984}. But shearing box radiation-MHD simulations by \citet{h2009a} suggested that the stress scales approximately with the total pressure. Simultaneously, the Lightman-Eardley viscous instability was also confirmed. The second method to eliminate the instability is to make the disk cooler, thus increasing the relative importance of gas pressure compared to radiation pressure. \citet{s1994} found that the disk would be stable if most of the gravitational energy released in the disk were transported to the corona. Convective cooling has been suggested as a stabilizing factor \citep{g1995}, although later research showed that it probably has a minor effect on disk stability. Turbulence, instead of convection, has also been suggested to play a key role in increasing the critical accretion rate \citep{z2013}. Another possible mechanism to cool the disk relies on magnetic pressure to provide part of the vertical hydrodynamical support \citep{z2011}. \citet{h2009b} pointed out that the time delay between the turbulent stress and total pressure of the disk can also make the disk stable. But \citet{j2013}, using the same code, found that the disk still runs away once they adopt a large enough horizontal shearing box size.

In this work, we investigate the thermal stability of a thin disk with winds. Strong winds driven by a large-scale magnetic field can take away most of the gravitational energy released in the disk, thereby reducing the disk temperature considerably \citep{l2012}. Thus, disk winds can help to cool the disk and make the disk stable. In this work we assume the existence of a large-scale magnetic field threading the disk; how this field is established remains an open question. The formation of large-scale field in a thin disk seems to be difficult due to its fast diffusive speed \citep{v1989,l1994}. \citet{c2013} suggested that the advection timescale can become smaller than the diffusion timescale in the presence of winds, thus the field can be effectively dragged inwards from the outer region even for a thin disk. We consider the advection and diffusion time-scales of the magnetic field based on this work in Section \ref{fields}.

\section{MODEL}\label{models}

We adopt the model of a relativistic thin disk
with magnetically driven outflows/jets around a Kerr black hole. The basic equations are basically the same as in \citet{l2012}, see
also \citet{a1996} and \citet{m2000}. The metric around the black
hole reads (geometrical units $G=c=1$ are adopted):
\begin{equation}
ds^{2}=-\frac{R^2 \Delta}{A} dt^2+\frac{A}{R^2}(d\phi-\omega
dt)^2+\frac{R^2}{\Delta}dR^2+dz^2, \label{metric}
\end{equation}
\begin{displaymath}
\Delta=R^2-2MR+a^2,
\end{displaymath}
\begin{displaymath}
A=R^4+R^2 a^2+2MRa^2,
\end{displaymath}
\begin{displaymath}
\omega=\frac{2MaR}{A},
\end{displaymath}
\begin{displaymath}
a=\frac{J}{M},
\end{displaymath}
where $M$ is the mass of the black hole, $J$ and $a$ are the angular
momentum and specific angular momentum of the black hole,
respectively, and $\omega$ is the dragging angular velocity of the
metric.

The steady state continuity equation is
\begin{equation}
\frac{d}{dR}(2\pi \Delta^{1/2} \Sigma v_{\rm R}/\gamma_\phi)+4\pi R \dot{m}_{\rm w}=0, \label{continuity}
\end{equation}
where $v_R$ is the radial velocity of the accretion flow and
$\Sigma=2\rho H$ is the surface density. Both viscous and magnetic torques can transfer the angular momentum of the accretion flow. In this work, we consider the contribution to the radial velocity from both of them. The Lorentz factor
$\gamma_\phi$ of the rotational velocity $v_\phi$ is given by
\begin{displaymath}
\gamma_\phi=(1-v_\phi^2)^{-1/2},
\end{displaymath}
\begin{displaymath}
v_\phi=A\tilde{\Omega}/R^2 \Delta^{1/2},
\end{displaymath}
and $\tilde{\Omega}=\Omega-\omega$, where $\Omega$ is the angular velocity.

The mass loss rate $\dot{m}_{\rm w}$ from unit surface area of the accretion disk can be obtained from
\begin{displaymath}
\dot{m}_{\rm w}=\frac{B_{\rm p}B_{\rm z}}{4\pi \Omega R} \mu
\label{massloss}
\end{displaymath}
\citep{c2013}, where $\mu$ is the dimensionless mass loading parameter of the outflow \citep{m1969}. The magnetic field is $B=(B_{\rm p}^2+B_{\phi}^2)^{1/2}$, where $B_{\phi}$ and $B_{\rm p}(=(B_{\rm R}^2+B_{\rm z}^2)^{1/2})$ are the toroidal and poloidal component of the fields, and $B_{\rm R}$ and $B_{\rm z}$ are the radial and vertical component of the fields, respectively. The inclination angle of field lines with respect to the mid-plane of the disk is required to be less than 60 degrees in order to launch jets from a Keplerian cold disk \citep{b1982}. We simply adopt 60 degrees in this work. The magnetic torque $T_{\rm m}\sim B_{\rm p} B_{\rm \phi}R/2\pi$ can also be written as $T_{\rm m}\sim 3R B_{\rm p}^2 \mu (1+\mu^{-2/3})/4\pi$ \citep*[using the cold approximation of Weber-Davis model, see][]{w1967,c2013}, implying that $\mu$ is $\sim 0.001$ for $B_{\rm \phi}=0.1 B_{\rm p}$. The mass loss rate is very small and can be neglected in this case. But when $B_{\rm \phi}=B_{\rm p}$, $\mu$ is $\sim 1$ and the mass loss rate is important. On the other hand, the mass loss rate is unimportant for $B_{\rm \phi}=10 B_{\rm p}$, although $\mu$ is $\sim 5$, because the poloidal fields decrease by an order of magnitude. Thus, the results for $B_{\rm \phi}=10 B_{\rm p}$ are qualitatively the same as those for $B_{\rm \phi}=0.1 B_{\rm p}$ (see Figs. \ref{f2}, \ref{f3} for details). We calculate the mass loss rate $\dot{m}_{\rm w}$ only when $B_{\rm \phi}=B_{\rm p}$ is adopted.

The radial momentum equation is
\begin{equation}
\frac{\gamma_\phi A M}{R^4 \Delta}\frac{(\Omega-\Omega_{\rm k}
^+)(\Omega-\Omega_{\rm k} ^-)}{\Omega_{\rm k} ^+ \Omega_{\rm k}
^-}+g_{\rm m}=0, \label{momentum}
\end{equation}
where we have neglected the radial pressure force. The Keplerian angular
velocities of the prograde ($+$) and retrograde ($-$) motions are
\begin{displaymath}
\Omega_{\rm k}^{\pm}=\pm\frac{M^{1/2}}{R^{3/2}\pm a M^{1/2}},
\end{displaymath}
and the radial magnetic force per unit mass is given by
\begin{displaymath}
g_{\rm m}={B_{\rm R}B_{\rm z}}/{2\pi\Sigma}.
\end{displaymath}

The angular momentum equation is
\begin{equation}
-\frac{\dot{M}}{2\pi} \frac{dL}{dR} + \frac{d}{dR}(R
W^R_\phi)+T_{\rm m}R=0, \label{angular}
\end{equation}
where the angular momentum of the accretion flow $L$ is
\begin{displaymath}
L=\frac{A^{1/2}(\gamma^{2}_{\phi}-1)^{1/2}}{R},
\end{displaymath}
and the height-integrated viscous tensor is
\begin{displaymath}
W^R_\phi=\alpha \frac{A^{3/2}\Delta^{1/2}\gamma _{\phi}^3}{R^6} W,
\end{displaymath}
where $\alpha$ is the Shakura-Sunyaev viscosity parameter. The
height-integrated pressure $W=2HP_{\rm tot}$, where the total
pressure $P_{\rm tot}=P_{\rm gas}+P_{\rm rad}+P_{\rm m}$; $P_{\rm
gas},P_{\rm rad}$ and $P_{\rm m}=B^2/8\pi$ are the gas pressure, radiation pressure and magnetic
pressure, respectively. The scale height $H$ of the
accretion disk is given by
\begin{displaymath}
H^2=c_{s}^2 R^4/(L^2-a^2),
\end{displaymath}
where $c_{\rm s}=\sqrt{P_{\rm tot}/\rho}$ is the sound speed of the
gas in the disk. The magnetic torque exerted on the accretion flow
due to the outflows/jets is
\begin{displaymath}
 T_{\rm m}={\frac {B_{\rm p} B_{\rm \varphi} R}{2\pi}}. \label{tm}
\end{displaymath}

The energy equation is
\begin{equation}
\nu \Sigma \frac{\gamma_{\phi}^4 A^2}{R^6}\left(
\frac{d\Omega}{dR}\right)^2= \frac{16acT^4}{3\bar{\kappa}\Sigma}.
\label{energy}
\end{equation}
where $\nu$ is the viscosity coefficient, $\nu \Sigma
d\Omega/dR=-\alpha W/R$ in $\alpha$-viscosity,  and $T$ is the
temperature of the gas at the disk midplane \citep{a1996}. The opacity
$\bar{\kappa}$ of the gas is given by
\begin{displaymath}
\bar{\rm \kappa}=\kappa_{\rm es}+\kappa_{\rm ff}=0.4+0.64\times
10^{23}\rho T^{-7/2} \textmd{cm}^2 \textmd{g}^{-1}, \label{opacity}
\end{displaymath}
where $\kappa_{\rm es}$ and $\kappa_{\rm ff}$ are the electron
scattering opacity and free-free opacity, respectively.

\section{RESULTS}\label{results}

\subsection{Analytical results for thin disk with winds}

In order to understand thin disk with winds better, we analyze the dynamical structure of thin disk using the approximations of \citet{s1973}. The basic equations are as follows in a Paczy$\acute{\rm n}$ski-Wiita potential \citep{p1980}:
\begin{equation}
\dot{M}=-2\pi R \Sigma V_{\rm R}, \label{continuity2}
\end{equation}

\begin{equation}
R(\Omega_{\rm k}^2-\Omega^2)-g_{\rm m}=0, \label{momentum2}
\end{equation}

\begin{equation}
-\frac{\dot{M}}{2\pi} \frac{d(\Omega R^2)}{dR} - \frac{d}{dR}(R^2
W_{\rm R\phi})+T_{\rm m}R=0, \label{angular2}
\end{equation}

\begin{equation}
\nu \Sigma {R^2}\left(
\frac{d\Omega}{dR}\right)^2= \frac{16acT^4}{3\bar{\kappa}\Sigma},
\label{energy2}
\end{equation}
where the Keplerian angular velocity $\Omega_{\rm k}^2=GM/R(R-R_{\rm g})^2$ and $W_{\rm R\phi}=2H\alpha P_{\rm tot}$ is the height-integrated viscous stress of the disk.

For simplicity, we adopt $B_{\rm \phi}=0.1 B_{\rm p}$ in this subsection in order to ignore the mass loss rate term in the continuity equation. Thus, the parameter $\beta_{\rm p}$ can be written as $\beta_{\rm p} \sim (P_{\rm gas}+P_{\rm rad})/(B^2/8\pi)$. In the momentum equation (\ref{momentum2}), the magnetic force $g_{\rm m}=B_{\rm R}B_{\rm z}/2\pi\Sigma <B^2/2\pi\Sigma=4P_{\rm tot}/(1+\beta_{\rm p})\Sigma=2\Omega_{\rm k}^2 H/(1+\beta_{\rm p})$, which is about $H/R$ smaller than $\Omega_{\rm k}^2 R$ and can be negligible even when $\beta_{\rm p}$ is far smaller than $1$. So the momentum equation can be rewritten as:
\begin{equation}
\Omega_{\rm k}\sim \Omega, \label{momentum3}
\end{equation}

The magnetic torque will dominate the transportation of gas angular momentum if the magnetic field is strong enough. With $\Omega_{\rm k}\sim \Omega$, the angular momentum equation (\ref{angular2}) can be rewritten as:
\begin{equation}
\dot{M}=\frac{4\pi T_{\rm m}}{\Omega}, \label{angular3}
\end{equation}
where the magnetic torque is $T_{\rm m}\sim B_{\rm p}B_{\rm \phi}R/2\pi \sim 0.1 B^2 R/2\pi\sim 0.4 R P_{\rm m}$ for $B_{\rm \phi}=0.1 B_{\rm p}$.

Using equation (\ref{momentum3}), energy equation (\ref{energy2}) can be rewritten:
\begin{equation}
\frac{16acT^4}{3\bar{\kappa}\Sigma}=\frac{9}{4}\nu\Sigma\Omega^2.
\label{energy3}
\end{equation}

\subsubsection{Gas pressure dominated outer disk}

We consider a gas pressure dominated outer disk with $P_{\rm gas}\gg P_{\rm rad}$ and $\sigma_{\rm T}\gg \sigma_{\rm ff}$ first. The opacity is $\bar{\kappa}=0.4 {\rm cm}^2 {\rm g}^{-1}$ and the total pressure is $P_{\rm tot}=(1+\beta_{\rm p})P_{\rm gas}/\beta_{\rm p}$, where $P_{\rm gas}=\rho k T/\mu m_{\rm p}$. With $\nu=\alpha c_{\rm s} H$, $P_{\rm tot}=\rho c_{\rm s}^2$ and $\Sigma=2\rho H$, the energy equation (\ref{energy3}) can be written as:
\begin{equation}
\frac{9}{4}\nu\Sigma\Omega^2=\frac{9}{4}\frac{(1+\beta_{\rm p})}{\beta_{\rm p}} \frac{\alpha k}{\mu m_{\rm p}} T \Sigma\Omega=\frac{16acT^4}{3\bar{\kappa}\Sigma}.
\end{equation}

Thus we get
\begin{equation}
T=\frac{3}{4} \left(\frac{1+\beta_{\rm p}}{\beta_{\rm p}}\right)^{1/3} \left(\frac{\alpha k \Omega \bar{\kappa}}{a c \mu m_{\rm p}}\right)^{1/3} \Sigma^{2/3}
\end{equation}
and the total pressure
\begin{equation}
P_{\rm tot}=\frac{(1+\beta_{\rm p})}{\beta_{\rm p}}\frac{\rho k T}{\mu m_{\rm p}}=\frac{\sqrt{3}}{4} \left(\frac{1+\beta_{\rm p}}{\beta_{\rm p}}\right)^{2/3}  \left(\frac{\alpha \bar{\kappa}}{ac}\right)^{1/6} \left(\frac{k}{\mu m_{\rm p}}\right)^{2/3} \Omega^{7/6} \Sigma^{4/3}.
\end{equation}
Assuming that the magnetic torque is responsible for all of the angular momentum transport, the mass accretion rate $\dot{M}$ is
\begin{displaymath}
\dot{M}=\frac{4\pi T_{\rm m}}{\Omega}=\frac{4\pi}{\Omega}\frac{0.4 R}{1+\beta_{\rm p}}P_{\rm tot}=\frac{2\sqrt{3}\pi}{5} \beta_{\rm p}^{-2/3} (1+\beta_{\rm p})^{-1/3} \left(\frac{\alpha \bar{\kappa}}{ac}\right)^{1/6} \left(\frac{k}{\mu m_{\rm p}}\right)^{2/3} \Omega^{1/6} R \Sigma^{4/3}
\end{displaymath}
\begin{equation}
=1.4*10^6 \beta_{\rm p}^{-2/3} (1+\beta_{\rm p})^{-1/3} \alpha^{1/6} \Omega^{1/6} R \Sigma^{4/3}.
\label{msigma}
\end{equation}

Consider a perturbation that slightly increases the surface density $\Sigma$ at radius $R$. The denser gas will at least partially concentrate the poloidal magnetic flux, leading to an increase of field strength. Thus, $\beta_{\rm p}$ must be a function of $\Sigma$ instead of a constant. We assume
\begin{equation}
B=B_0 \left(\frac{\Sigma}{\Sigma_0}\right)^{\epsilon}
\label{bsigma}
\end{equation}
in this work, where $B_0$ and $\Sigma_0$  are the initial field strength and surface density, respectively, and  $0<\epsilon<1$ is adopted. The parameter $\epsilon$ is defined as
\begin{equation}
\epsilon=\frac{\tau_{\rm dif}/\tau_{\rm adv}}{1+\tau_{\rm dif}/\tau_{\rm adv}}=\frac{\kappa_0 | v_{\rm R}|/\alpha c_{\rm s}}{1+\kappa_0 | v_{\rm R}|/\alpha c_{\rm s}},
\label{espilon}
\end{equation}
where $\tau_{\rm dif}$ and $\tau_{\rm adv}$ are the diffusion and advection timescales of the field, respectively, and $\kappa_0 =B_{\rm z}/B_{\rm{r,s}}$ is the inclination of field to the horizontal plane (see section \ref{fields} for details). Here the radial velocity $v_R$ comes from both the viscous and magnetic torques. When the diffusion timescale is far smaller than the advection timescale of the field ($\tau_{\rm dif} \ll \tau_{\rm adv}$), the field strength will be a constant and $\epsilon=0$. But if $\tau_{\rm dif} \gg \tau_{\rm adv}$, all the flux will be advected inwards effectively and the magnetic flux ($\Phi=B/ \Sigma$) will be a constant, which corresponds to $\epsilon=1$. Such a case would apply, for example, if the magnetic torque were entirely responsible for the radial velocity. If $\beta_{\rm p}\gg 1$, $\beta_{\rm p}$ is given by
\begin{equation}
\beta_{\rm p}\simeq \frac{P_{\rm tot}}{B^2/8\pi}=8\pi P_{\rm tot} B_0^{-2} \Sigma_0^{2\epsilon} \Sigma^{-2\epsilon}.
\label{betap}
\end{equation}

Combining with equation (\ref{msigma}), the resulting mass accretion rate (ignoring viscous stresses) can be written as
\begin{equation}
\dot{M}=\frac{4\pi}{\Omega}\frac{0.4 R}{1+\beta_{\rm p}}P_{\rm tot}\simeq\frac{0.2 B_0^2 \Sigma_0^{-2\epsilon}  R}{\Omega} \Sigma^{2 \epsilon}
\label{msigma2}
\end{equation}
when $\beta_{\rm p}\gg 1$.


\subsubsection{Radiation pressure dominated inner disk}

For a radiation pressure dominated inner disk, the radiation pressure $P_{\rm rad}\gg P_{\rm gas}$ and $\sigma_{\rm T}\gg \sigma_{\rm ff}$. The total pressure $P_{\rm tot}=(1+\beta_{\rm p})P_{\rm rad}/\beta_{\rm p}$, where $P_{\rm rad}=a T^4/3$. The energy equation (\ref{energy3}) can be written as:
\begin{equation}
\frac{9}{4}\nu\Sigma\Omega^2=\frac{\alpha (1+\beta_{\rm p})^2 a^2 T^8}{\beta_{\rm p}^2 \Sigma\Omega }=\frac{16acT^4}{3\bar{\kappa}\Sigma}.
\end{equation}
Thus the temperature and the total pressure are given by
\begin{equation}
T^4=\frac{16 \beta_{\rm p}^2 \Omega c } {3\alpha \bar{\kappa} (1+\beta_{\rm p})^2 a}
\end{equation}
and
\begin{equation}
P_{\rm tot}=\frac{16}{9}\frac{\beta_{\rm p} \Omega c}{\alpha \bar{\kappa} (1+\beta_{\rm p})}.
\end{equation}

The mass accretion rate is
\begin{equation}
\dot{M}=\frac{4\pi}{\Omega}T_{\rm m}=\frac{1.6 \pi R}{\Omega (1+\beta_{\rm p})}P_{\rm tot}=\frac{8.9 \beta_{\rm p} c R}{\alpha \bar{\kappa} (1+\beta_{\rm p})^2}.
\end{equation}

\begin{figure}
\includegraphics[width=15cm]{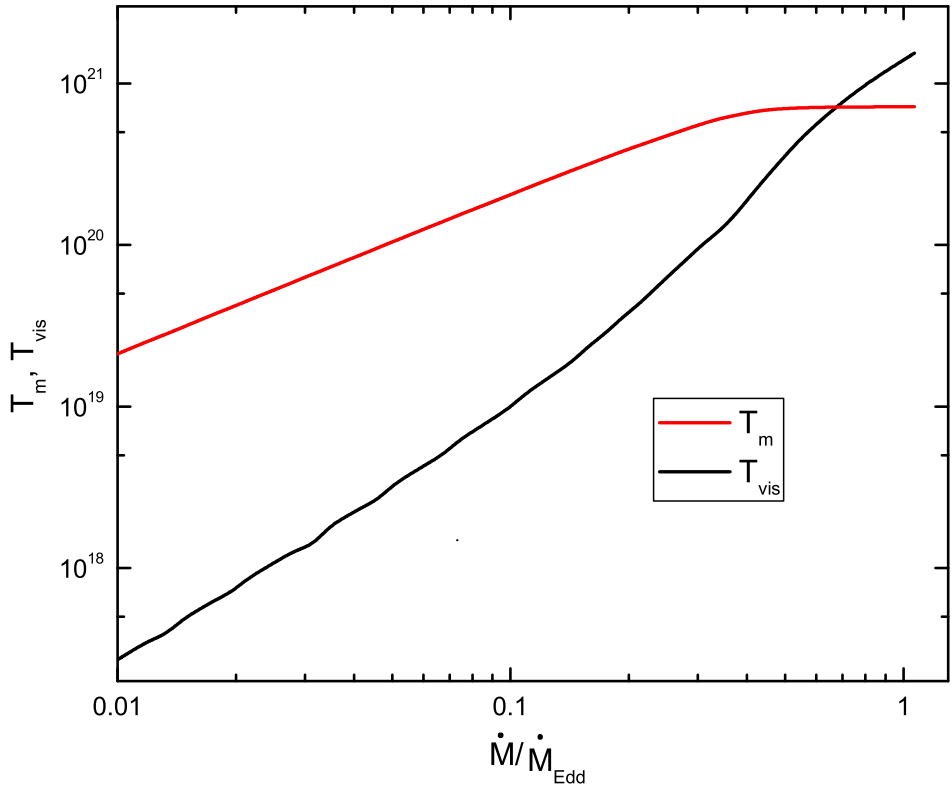}
\caption{The viscous and magnetic torques as functions of mass accretion rate at $R=2R_{\rm ISCO}$, where $M=10M_{\odot}$, $a=0.9$, $\alpha=0.1$, $B_{\rm \phi}=0.1 B_{\rm p}$ and $\beta_{\rm p}=10$ are adopted. \label{f1}}
\end{figure}

Using equation (\ref{betap}), we can get the same $\dot{M}-\Sigma$ equation as equation (\ref{msigma2}) when $\beta_{\rm p}\gg 1$. It seems that the whole disk should be very stable because both the gas and radiation dominated disk regions share the same positive slope $2\epsilon$. But actually, this solution of a radiation dominated inner disk with inflow driven by magnetic torque is hard to realize for the reason that the viscous torque ($T_{\rm vis}\sim P_{\rm tot} 2H\alpha $) is comparable to or even larger than the magnetic torque [$T_{\rm m} \sim P_{\rm tot} 0.4 R/(1+\beta_{\rm p})$] when the accretion rate is close to the Eddington accretion rate (Fig. \ref{f1}). So the premise of magnetic torque dominating is no longer correct. The real case is that both viscous and magnetic torques will exist in the disk and the slope of the $\dot{M}-\Sigma$ curve in the radiation pressure dominated region is between $2\epsilon$ (magnetic torque dominated) and $-1$ (no magnetic field) (see Figs. \ref{f2}, \ref{f3}).


\subsection{Numerical results for thin disk with winds}

\subsubsection{Numerical methods}

We numerically solve Equations
(\ref{continuity})$-$(\ref{energy}) in this work. The continuity equation (\ref{continuity})
and the angular momentum equation (\ref{angular}) can be rewritten as

\begin{equation}
(2\pi \Delta^{1/2} \Sigma v_{\rm R}/\gamma_\phi)\mid ^{\rm {R+\Delta R}}_{\rm {R}}+ (4\pi R \dot{m}_{\rm w})\mid_{\rm
{R+\Delta R}} \Delta R=0 \label{continuity4}
\end{equation}
and
\begin{equation}
-\frac{\dot{M}}{2\pi}L\mid ^{\rm {R+\Delta R}}_{\rm {R}}+
(RW^R_\phi)\mid ^{\rm {R+\Delta R}}_{\rm {R}}+ (T_{\rm m}R)\mid_{\rm
{R+\Delta R}} \Delta R=0, \label{angular4}
\end{equation}
respectively, when $\Delta R \rightarrow 0$. The inner radius of the accretion
disk $R_{\rm in}$ is set at the innermost stable circular orbit
(ISCO), where the zero viscous torque condition $W^R_\phi \mid _{\rm
{R_{\rm ISCO}}}=0$ is adopted. $(2\pi \Delta^{1/2} \Sigma v_{\rm R}/\gamma_\phi)\mid _{\rm {R_{ISCO}}}$ is the mass accretion rate at the ISCO and the angular momentum at the ISCO is set
to $L \mid _{\rm {R_{\rm ISCO}}}=L_{\rm k} \mid _{\rm
{R_{\rm ISCO}}}$, where $L_{\rm k}$ is the Keplerian angular momentum, which is a very good approximation (see equation
\ref{momentum3}).

Combining equations (\ref{momentum}), (\ref{energy}), (\ref{continuity4}) and (\ref{angular4}),
with the parameters $M, \dot{M}, a, \beta_{\rm p}$ and the Shakura-Sunyaev
parameter $\alpha$, the four variables $\rho$, $v_{\rm R}$, $\Omega$ and $T$
can be numerically solved by the Newton-Raphson method for nonlinear
equations. At first, we calculate the disk properties at $R=R_{\rm
ISCO}+ \Delta R$ as the values of $W^R_\phi \mid _{\rm {R_{\rm
ISCO}}}$, $(2\pi \Delta^{1/2} \Sigma v_{\rm R}/\gamma_\phi)\mid _{\rm {R_{ISCO}}}$ and $L \mid _{\rm {R_{\rm ISCO}}}$ are known. With the disk
properties at $R$, the disk structure at $R+\Delta R$ can be gotten too.
Similarly, we can gradually obtain the properties of the whole
accretion disk from $R_{\rm ISCO}$ to the outer radius $R_{\rm
out} (=1000 R_{\rm ISCO})$.

\subsubsection{Results}

\begin{figure}
\includegraphics[width=16cm]{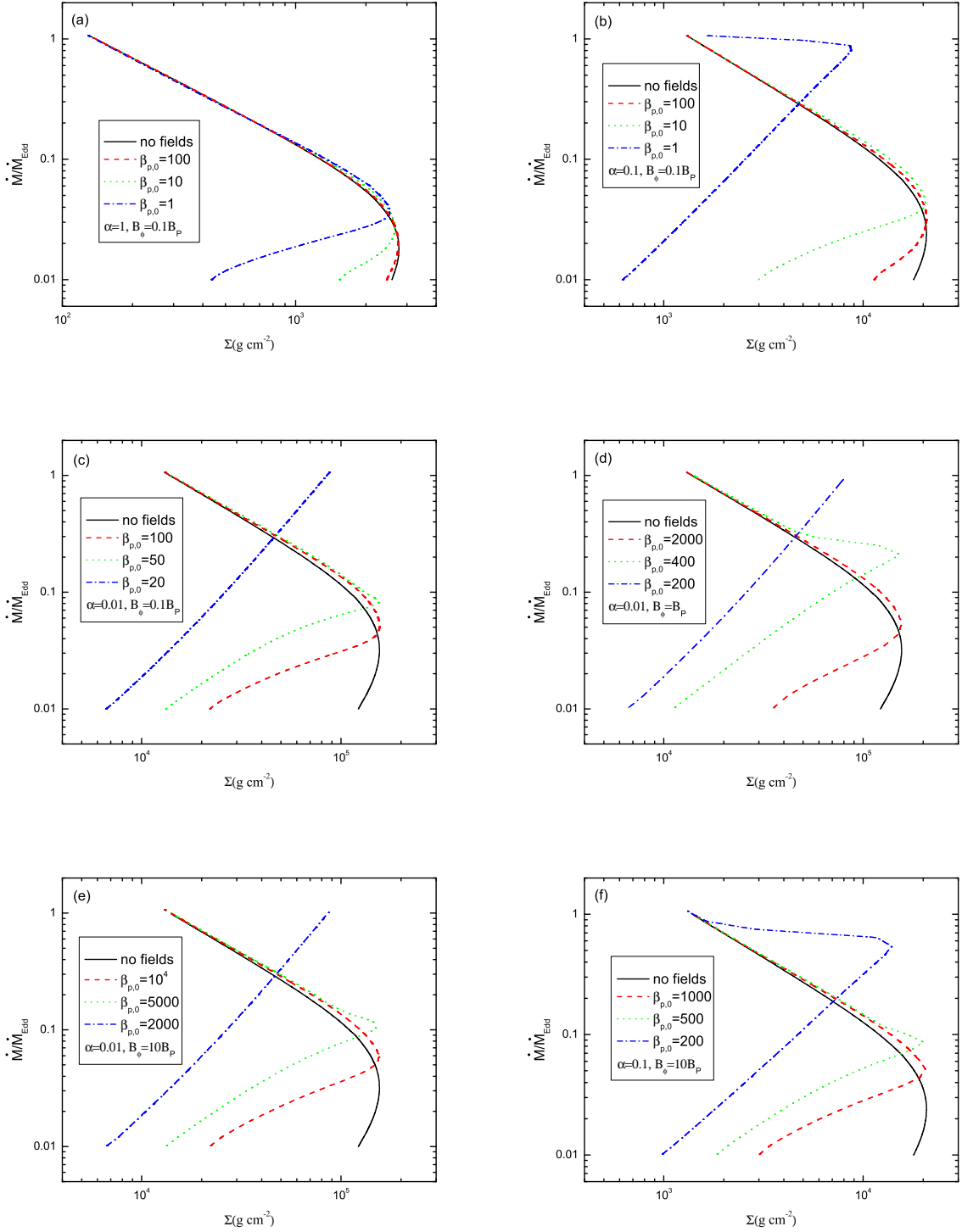}
\caption{The $\dot{M}-\Sigma$ curve of a thin disk with winds at radius $R=2R_{\rm ISCO}$, where $M=10M_{\odot}$ and $a=0.9$ are adopted.  \label{f2}}
\end{figure}

\begin{figure}
\includegraphics[width=16cm]{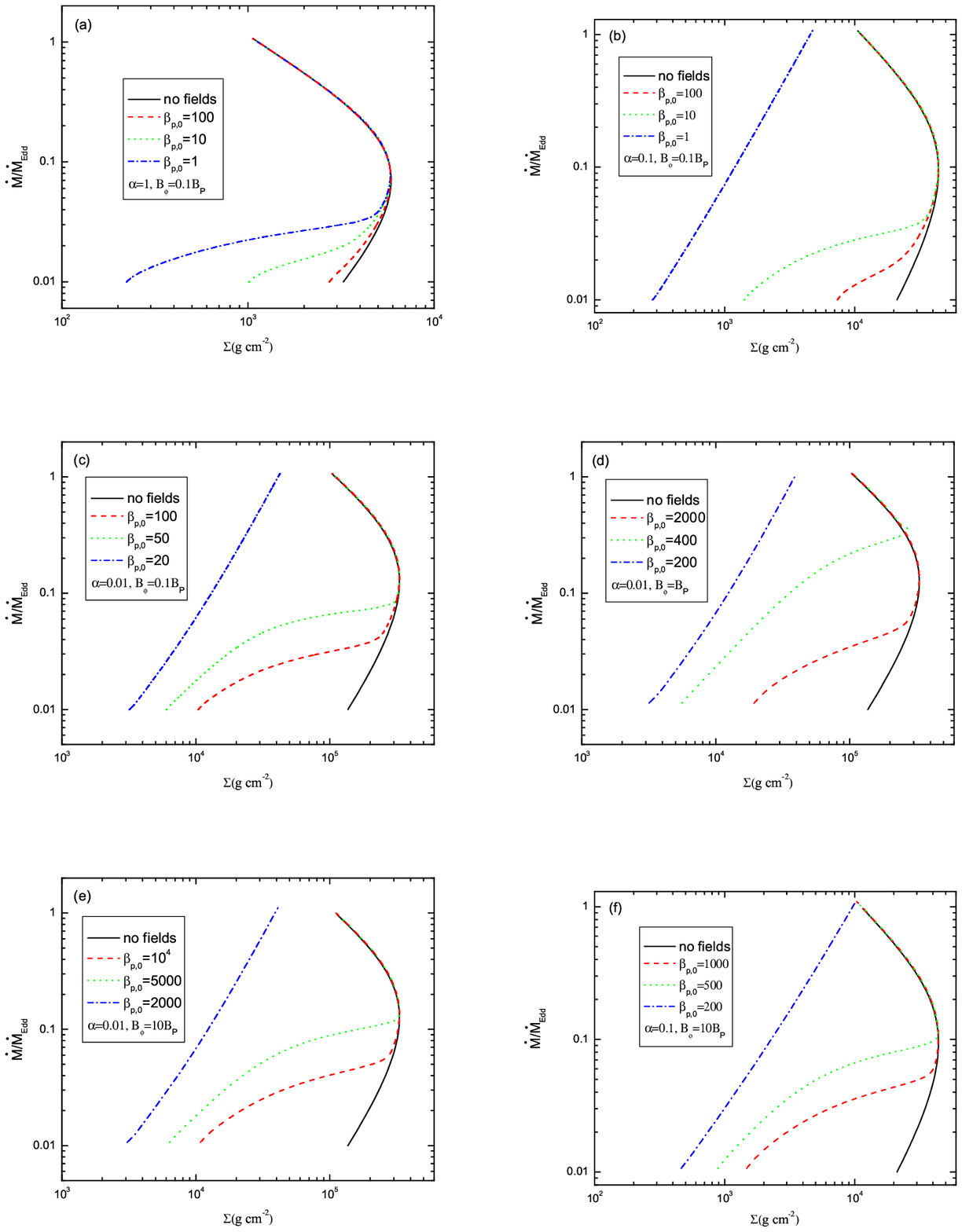}
\caption{The same as Fig. \ref{f2} except that the radius $R=10R_{\rm ISCO}$. \label{f3}}
\end{figure}

\begin{figure}
\includegraphics[width=15cm]{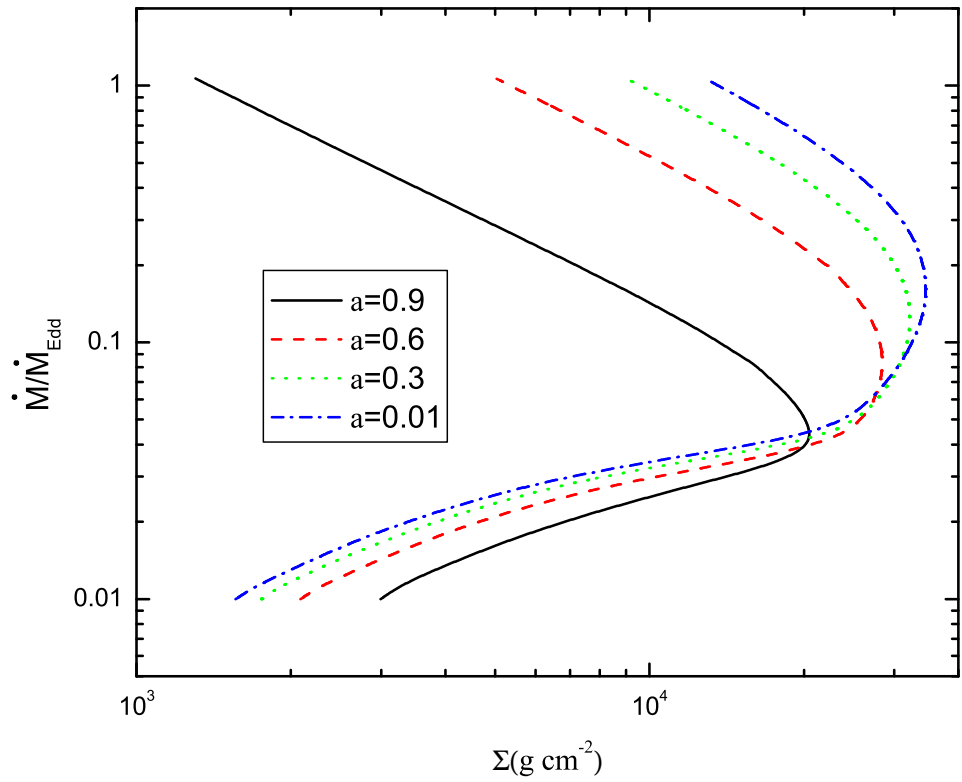}
\caption{The $\dot{M}-\Sigma$ curve of a thin disk with winds at radius $R=2R_{\rm ISCO}$, where $\alpha=0.1$, $B_{\rm \phi}=0.1 B_{\rm p}$ and $\beta_{\rm p,0}=10$ are adopted. \label{f4}}
\end{figure}

\begin{figure}
\includegraphics[width=15cm]{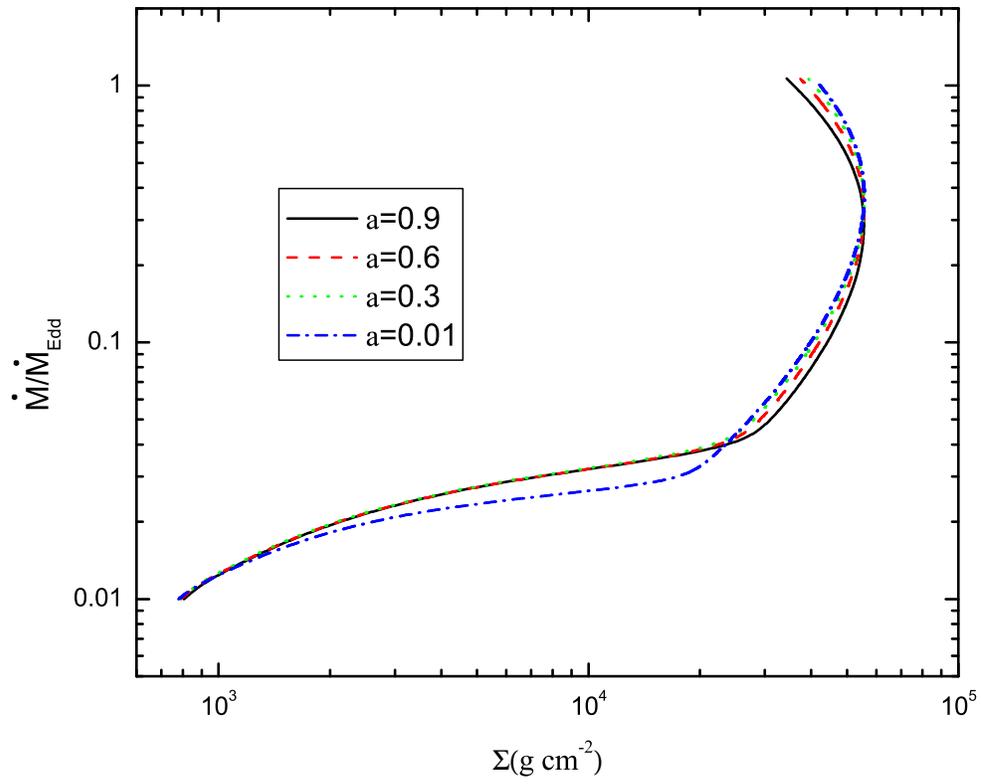}
\caption{The same as Fig. \ref{f4} except that the radius $R=60R_{\rm g}$. \label{f5}}
\end{figure}

The effects of a disk wind on the thermal stability of the disk are studied through the $\dot{M}-\Sigma$ curves for various parameters in Figs. \ref{f2} $-$ \ref{f5}, where the black hole mass $M=10 M_{\odot}$ is always adopted. All the calculations start from $\dot{M}/\dot{M}_{\rm Edd}=0.01$ ($\dot{M}_{\rm Edd}=1.5\times10^{18}M/M_{\odot}gs^{-1}$). In order to get the values of $B_0$ and $\Sigma_0$ in equation (\ref{espilon}), we adopt $\beta_{\rm p}=\beta_{\rm p,0}$ when $\dot{M}/\dot{M}_{\rm Edd}=0.01$.


In all the $\dot{M}-\Sigma$ figures, the negative slope always represents that the disk is radiation pressure dominated, which is both thermally and viscously unstable. The transition point from positive to negative slope corresponds to the location of the critical mass accretion rate $\dot{M}_{\rm crit}$. The wind can take away lots of energy and make the disk cooler. But once the radiation pressure dominates the gas pressure, the slope will change sign. All the $\dot{M}-\Sigma$ curves are plotted at radius $R=2 R_{\rm ISCO}$ in Fig. \ref{f2}. It is found that, with the presence of the disk wind, the critical accretion rate corresponding to the thermal instability can be significantly increased. The disk is quite stable even for a very weak initial poloidal magnetic field ($\beta_{\rm p,0}\sim 2000$) for $\alpha=0.01$ and $B_{\rm \phi}=10B_{\rm _p}$. But a somewhat stronger (but still weak) field ($\beta_{\rm p,0}\sim 200$ or $\beta_{\rm p,0}\sim 20$) is required to make the disk stable when $B_{\rm \phi}=B_{\rm _p}$ or $B_{\rm \phi}=0.1B_{\rm _p}$ is adopted. The slope for a gas pressure dominated disk is found to be steeper when the field is stronger (which means larger $\epsilon$), which is roughly consistent with our analytical results ($\dot{M}\sim\Sigma^{2\epsilon}$). We consider $0.01\leq\alpha\leq1$ as suggested by MHD simulations \citep[e.g.,][]{b2013}. The critical accretion rate is found to have a negative relation with $\alpha$ (see Figs. \ref{f2}, \ref{f3}). Smaller $\alpha$ corresponds to a larger critical accretion rate for the same initial field strength $\beta_{\rm p,0}$. Since most of the gravitational energy is dissipated within the region $R\leq 10 R_{\rm ISCO}$, we give the analogous results at $R=10 R_{\rm ISCO}$ in Fig. \ref{f3}, which are very similar to those of Fig. \ref{f2} but showing greater stability. As we suggested in section \ref{models}, the results for $B_{\rm \phi}=10B_{\rm _p}$ and $B_{\rm \phi}=0.1 B_{\rm _p}$ are similar to each other (see, e.g., Figs. \ref{f2}c and \ref{f2}e).

The effect of black hole spin $a$ is investigated in Figs. \ref{f4} and \ref{f5}. It is found that the accretion disk around a rapidly spinning black hole has a lower instability threshold at $R=2 R_{\rm ISCO}$ (Fig. \ref{f4}). Because $R_{\rm ISCO}$ varies with spin $a$, we study the effect of spin at constant radius $R=60 R_{\rm g}$ ($\sim 10 R_{\rm ISCO}$ when $a=0.01, R_{\rm g}=GM/c^2$) in Fig. \ref{f5} and find that at such large radii the spin has little effect. The disk is more stable when $\alpha$ and spin $a$ are smaller, which is qualitatively the same as the results of \citet{z2013}.

\section{THE FORMATION OF LARGE-SCALE MAGNETIC FIELD} \label{fields}

The large-scale magnetic field threading the accretion disk plays a key role in the formation of winds and jets. However, how the large-scale field can be constructed is still an open issue. A promising way is that there is large-scale magnetic field at the outer boundary of the disk and the field may be dragged into the inner disk through the accretion of gas.  Whether or not the field can be effectively advected depends on the balance between the advection timescale $\tau_{\rm adv}$ and the diffusion timescale $\tau_{\rm dif}$ of the field. For a geometrically thick disk ($H\sim R$), the large-scale field can probably be advected inwards due to its fast radial advection velocity \citep{l1994,c2011}. But the diffusive process dominates for a standard thin disk ($H << R$) so the advection of the field may be very ineffective \citep{v1989,l1994}.

There are several factors that may affect the formation of the large-scale field in a thin disk. For example, the diffusive process will be suppressed by the presence of strong magnetic field \citep{s2005} or the advection process can be accelerated by the external torque induced by the wind. \citet{c2013} found that even for a moderately weak field, the wind can significantly improve the efficiency of advection by taking away angular momentum from the disk, which results in the increase of the radial velocity. In this section, we extend the research of \citet{c2013} by studying the timescales of advection and diffusion in the whole disk.

The advection and diffusion timescales are

\begin{equation}
\tau_{\rm {adv}}\sim \frac{R}{|v_{\rm R}|}
\label{advecton}
\end{equation}
and
\begin{equation}
\tau_{\rm {dif}}\sim \frac{RH\kappa_0}{\eta},
\label{diffusion}
\end{equation}
respectively \citep{c2013}, where $\eta$ is magnetic diffusivity. $\tau_{\rm {dif}}$ is calculated with $\beta_{\rm p}= \infty$ when the magnetic field is absent. According to recent MHD simulations, the magnetic Prandtl number $Pr_{\rm m}=\eta/\nu$ is always $\sim 1$ \citep*[e.g.,][]{f2009,g2009}. In this work, we simply adopt $\eta \sim \nu$ for all the calculations.

\begin{figure}
\includegraphics[width=15cm]{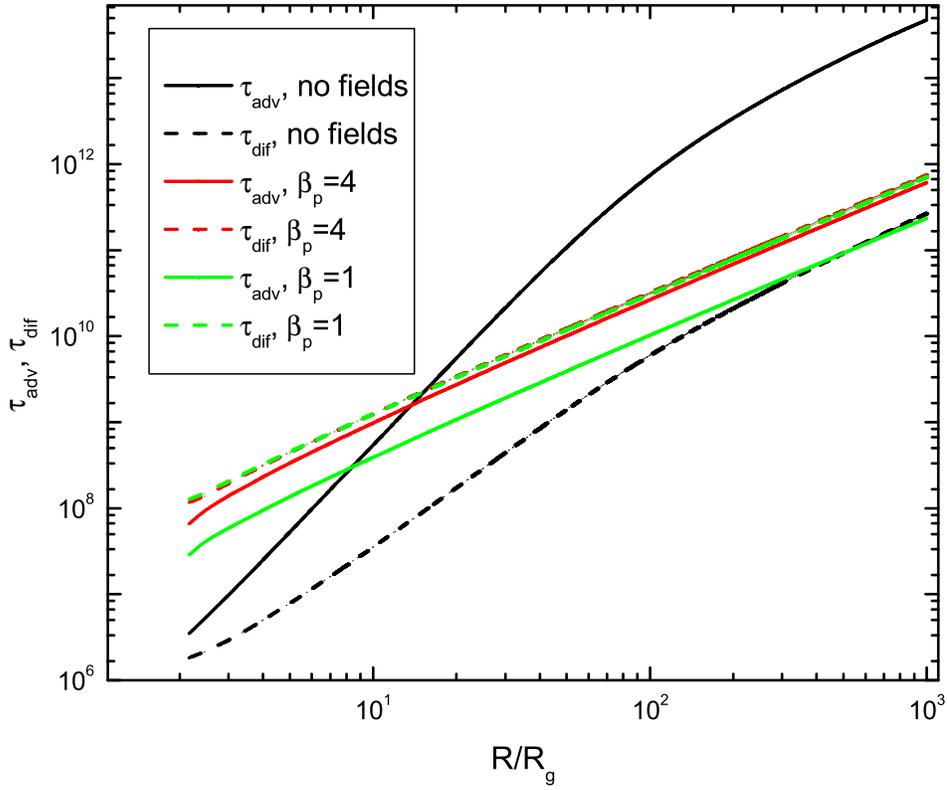}
\caption{The advection timescale $\tau_{\rm adv}$ and the diffusion timescale $\tau_{\rm dif}$ as functions of radius, where $M=10^8 M_{\odot}$, $\kappa_{0}=\sqrt{3}$, $\alpha=0.1$, $B_{\rm \phi}=0.1 B_{\rm p}$ and $\dot{M}/\dot{M}_{\rm {Edd}}=0.1$ are adopted. \label{f6}}
\end{figure}

\begin{figure}
\includegraphics[width=15cm]{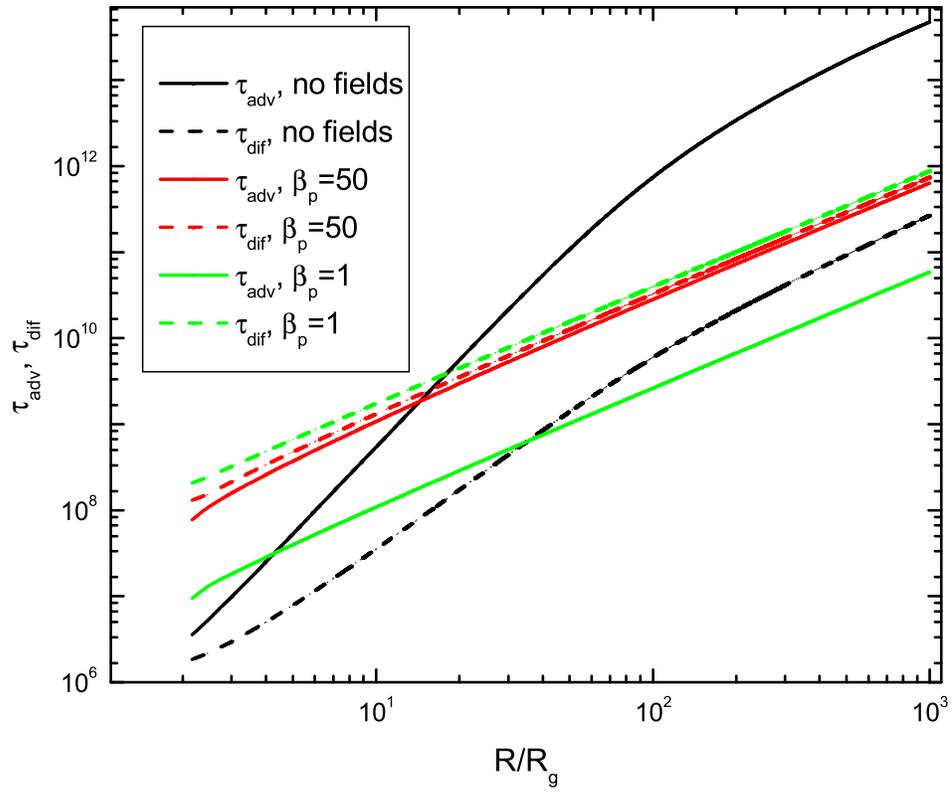}
\caption{The same as Fig. \ref{f6} except that $B_{\rm \phi}=B_{\rm p}$. \label{f7}}
\end{figure}

\begin{figure}
\includegraphics[width=15cm]{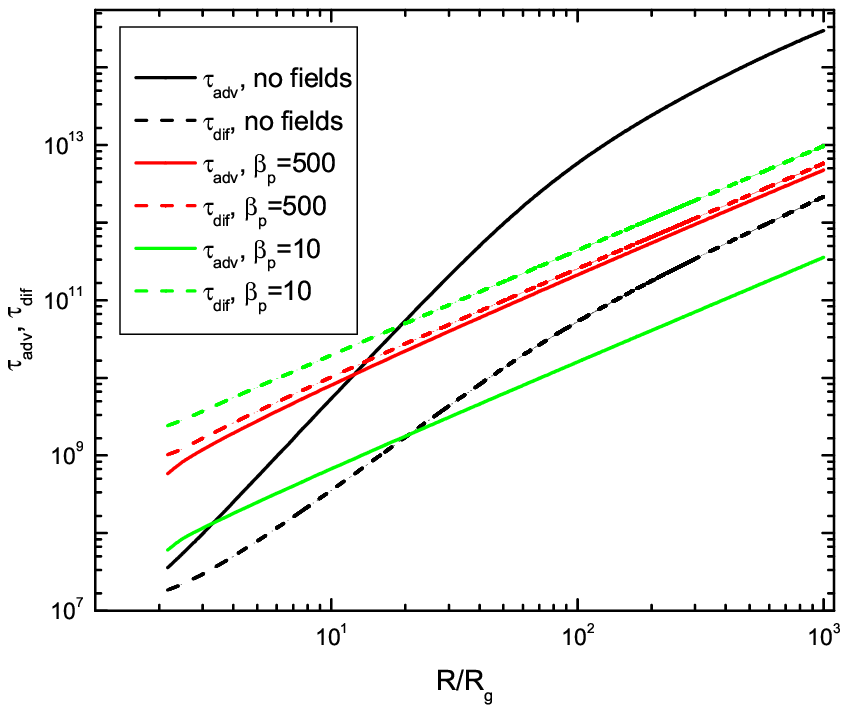}
\caption{The same as Fig. \ref{f6} except that $\alpha=0.01$ and $B_{\rm \phi}=B_{\rm p}$. \label{f8}}
\end{figure}

\begin{figure}
\includegraphics[width=15cm]{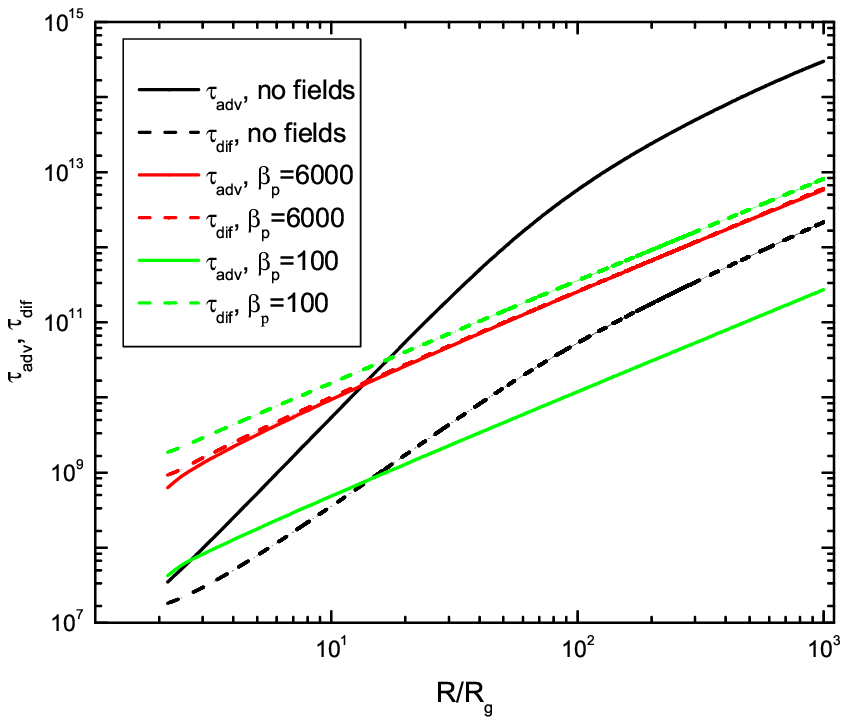}
\caption{The same as Fig. \ref{f6} except that $\alpha=0.01$ and $B_{\rm \phi}=10 B_{\rm p}$. \label{f9}}
\end{figure}

Our results are basically the same as \citet{c2013}, i.e., the advection timescale $\tau_{\rm adv}$ can be smaller than the diffusion timescale $\tau_{\rm dif}$ if there are strong enough winds driven by magnetic field (Fig. \ref{f6}). But in contrast to the results of \citet{c2013}, the main reason for $\tau_{\rm adv}< \tau_{\rm dif}$ is that the diffusion timescale increases a lot due to the decrease of viscosity $\nu$ ($\nu \sim \eta$), which is induced by the decrease of gas temperature resulting from the enormous energy taken away by the winds. The advection timescale $\tau_{\rm adv}$ does decrease in the middle and outer regions of a thin disk due to the increasing radial velocity. But in the inner disk region, the advection timescale of a disk with winds is found to be even larger than that of a standard disk when the radiation dominated inner disk disappears. The critical value of the magnetic field strength is $\beta_{\rm p} \leq 4$ in order to satisfy $\tau_{\rm adv}\leq \tau_{\rm dif}$ for $\alpha=0.1$ and $B_{\rm \phi}=0.1 B_{\rm _p}$. And $\tau_{\rm adv}$ becomes smaller compared with $\tau_{\rm dif}$ for smaller $\beta_{\rm p}$. In Figs. \ref{f7} $-$ \ref{f9}, we show that the critical value of magnetic field parameter $\beta_{\rm p}$ for $\tau_{\rm adv}< \tau_{\rm dif}$ increases from $4$ to $50$, $500$ and $6000$ for $\alpha=0.1$, $B_{\rm \phi}=B_{\rm _p}$; $\alpha=0.01$, $B_{\rm \phi}=B_{\rm _p}$ and $\alpha=0.01$, $B_{\rm \phi}=10B_{\rm _p}$, respectively. The field can be effectively dragged inwards for a very weak poloidal field ($\beta_{\rm p}\sim 100$) when $\alpha=0.01$ and $B_{\rm \phi}=10B_{\rm _p}$ are adopted. The mass loss rate $\dot{m}_{\rm w}$ is included when $B_{\rm \phi}=B_{\rm _p}$ for all the figures. The largest total mass loss rate is about $30\%$ of the mass accretion rate at the outer boundary.

\section{CONCLUSIONS AND DISCUSSION} \label{conclusionS}

\begin{figure}
\includegraphics[width=15cm]{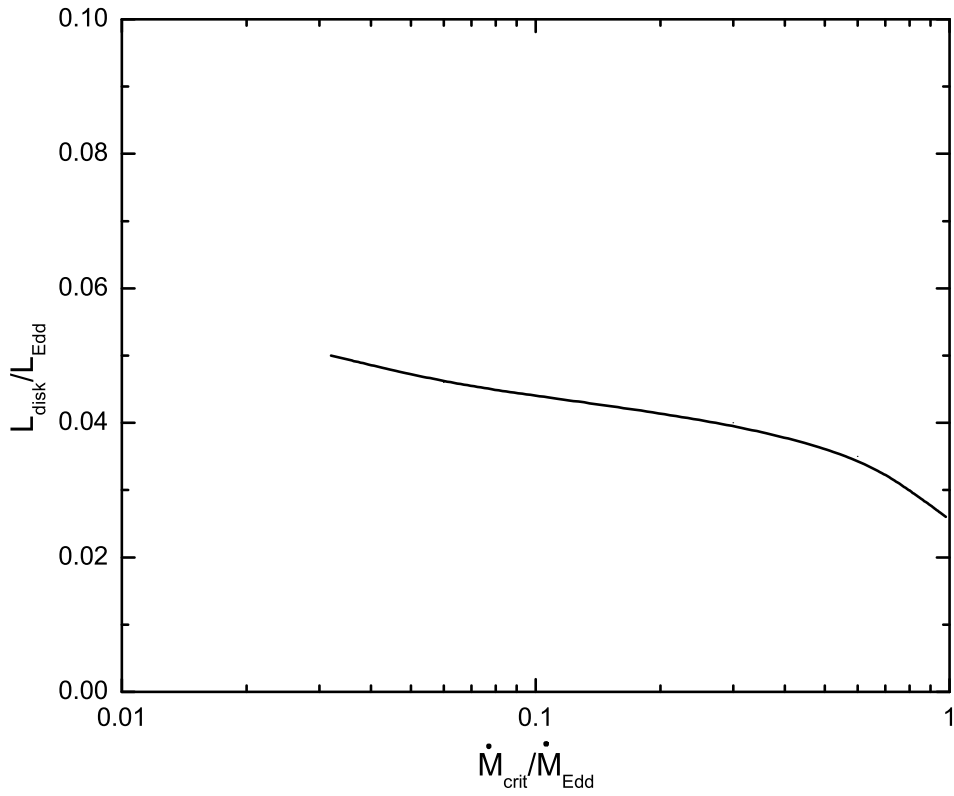}
\caption{Eddington scaled disk luminosity as a function of $\dot{M}_{\rm crit}/\dot{M}_{\rm Edd}$, where $M=10M_{\odot}$, $\alpha=0.01$ and $B_{\rm \phi}=10 B_{\rm p}$ are adopted. \label{luminosity}}
\end{figure}

In this work, we investigate the $\dot{M}-\Sigma$ curves of a thin disk with magnetically driven winds. It is found that, because disk winds can greatly decrease the disk temperature, the critical accretion rate $\dot{M}_{\rm crit}$ can be increased significantly and the disk becomes more stable (Figs. \ref{f2}, \ref{f3}). It seems that both the gas and radiation pressure dominated regions possess the same slopes in the $\dot{M}-\Sigma$ curves ($\dot{M}\sim\Sigma^{2\epsilon}$) when magnetic torques drive the inflow. But the real slope in the radiation pressure dominated region is between $-1$ and $2\epsilon$ in numerical calculations because both the viscous torque and magnetic torque are important when the mass accretion rate is close to the Eddington accretion rate. The parameter $\alpha$, the strength and the morphology of the initial magnetic fields all strongly affect the critical accretion rate $\dot{M}_{\rm crit}$. If $\beta_{\rm p,0}$ and $\alpha$ are smaller, the thin disk will be more stable. While the accretion disk with winds becomes stable for a high accretion rate, the luminosity threshold may not increase because of the much lower gas temperature. Indeed, it is found that the disk luminosity corresponding to $\dot{M}_{\rm crit}$ remains almost constant or even decreases slowly with the increase of $\dot{M}_{\rm crit}$ (Fig. \ref{luminosity}). Thus the absence of thermal instability in luminous accretion systems is still a problem even if the disk is stable for a very high accretion rate, unless other components, such as a corona or winds, contribute significantly to the luminosity in relevant spectral bands.

\begin{figure}
\includegraphics[width=15cm]{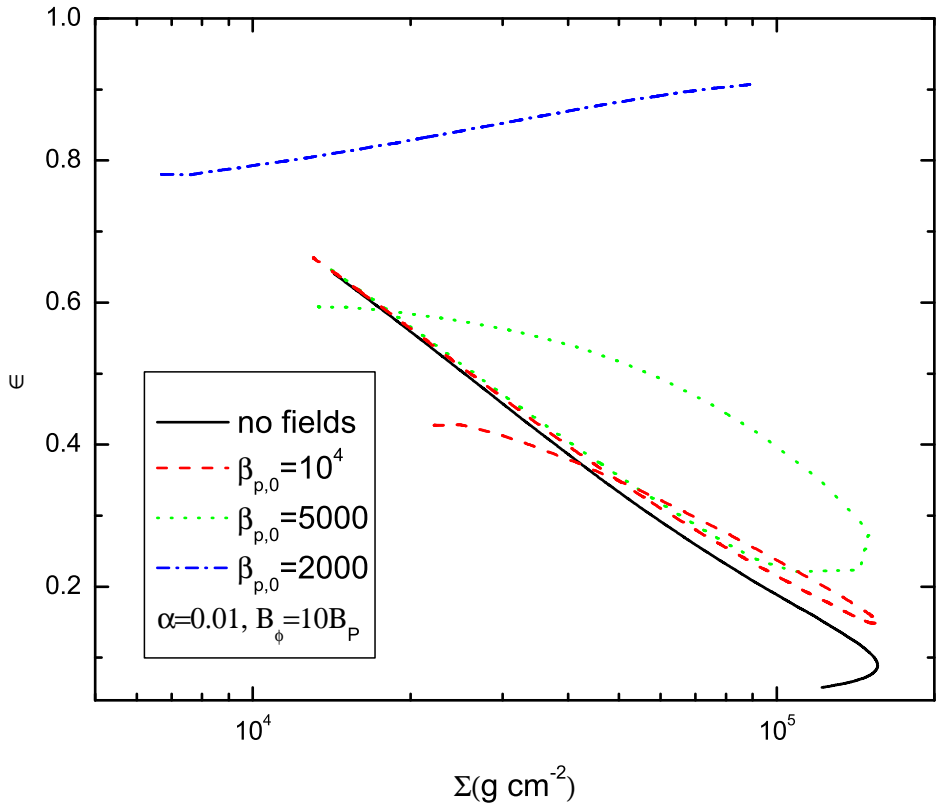}
\caption{$\epsilon$ as functions of surface density $\Sigma$ at $R=2R_{\rm ISCO}$, corresponding to Fig. \ref{f2}e. \label{f10}}
\end{figure}

Using equation (\ref{betap}) ($\beta_{\rm p}\sim \Sigma^{4/3-2\epsilon}$), it is interesting to note that there is a critical initial field strength $\beta_{\rm p,0,crit}$ corresponding to $\epsilon=2/3$ ($\beta_{\rm p}\sim \Sigma^{0}$). If $\beta_{\rm p,0} < \beta_{\rm p,0,crit}$, $\beta_{\rm p}$ will become smaller and smaller with increasing $\Sigma$ and the disk will tend to be more stable. Otherwise, $\beta_{\rm p}$ will become larger and the disk will be like the standard thin disk with increasing $\Sigma$ if $\beta_{\rm p,0} > \beta_{\rm p,0,crit}$. Thus, the slopes in the $\dot{M}-\Sigma$ curves will tend to be  either larger or smaller with increasing $\Sigma$ (see Figs. \ref{f2}, \ref{f3}). We have studied how $\epsilon$ varies with the surface density $\Sigma$ for different initial poloidal fields at $R=2R_{\rm ISCO}$, for example, in Fig. \ref{f10}, where $\alpha=0.01$ and $B_{\phi}=10B_{\rm p}$ are adopted. Only the dash-dotted line ($\beta_{\rm p,0}=2000$) satisfies $\beta_{\rm p,0} < \beta_{\rm p,0,crit}$ for the surface densities considered. Furthermore, both the field strength and $\epsilon$ increase with increasing $\Sigma$. On the contrary, $\beta_{\rm p}$ becomes larger with increasing $\Sigma$ for both the dashed and dotted lines at first, because $\beta_{\rm p,0} > \beta_{\rm p,0,crit}$. As a result, the magnetic field will be unimportant and $\epsilon$ is almost the same as that of a standard disk (the black line).

While the disk seems to be quite stable in the presence of winds, there are still two major open questions: a) the presence of the winds driven by large-scale magnetic field; and b) the formation of a large-scale field. This model predicts strong winds driven by the field, which seem to be absent in most X-ray binaries. But in order to make the winds visible, observationally, they would have to interact with ambient gas or produce internal shocks. So if there aren't internal shocks in the winds, the winds will be invisible due to the lack of ambient gas surrounding the disk in X-ray binaries. But X-ray absorption lines resulting from disk winds may be detected when the mass loss rate is important ($B_{\rm \phi}\sim B_{\rm p}$). Such absorption lines do seem to be present in some X-ray binaries \citep{k2012,m2012b}.

The formation of a large-scale field threading a thin disk is a key point in this work. We investigate the advection and diffusion timescales of the field in the disk in section \ref{fields}. Our results are basically the same as those of \citet{c2013}, i.e., the advection timescale can be smaller than the diffusion timescale and the field can be effectively dragged inwards, if the field is initially strong enough. But the main reason for this is that the wind takes away lots of the gravitational energy and so the diffusion timescale becomes larger (Figs. \ref{f6}$-$\ref{f9}). However, even if the field can be effectively dragged inwards, the formation of the large-scale field still depends on the outer boundary conditions. An original large-scale field is needed on the outer boundary of the disk. MHD simulations also suggest the formation of large-scale field depending on the outer boundary conditions \citep{b2008,m2012}. But where the original field comes from is still an unsolved problem.

\section* {ACKNOWLEDGEMENTS}
We thank the referee for his/her very thorough and helpful report. This work is supported by the NSFC (grants 11233006, 11373056) and the Science and Technology Commission of Shanghai Municipality (13ZR1447000).

\end{document}